\documentclass[namedreferences]{solarphysics}

\usepackage[urlcolor=blue,colorlinks=true,linkcolor=red,citecolor=magenta,breaklinks=true]{hyperref}
\usepackage[optionalrh,natbib]{spr-sola-addons} % For Solar Physics
\usepackage{graphicx}        % For eps figures, newer & more powerfull
\usepackage{color}           % For color text: \color command
\usepackage{breakurl}        % For breaking URLs easily trough lines
\usepackage{rotating}        % For rotate figure + caption
\usepackage{longtable}
\usepackage{enumerate}
\graphicspath{{figs/}}

\newcommand{\adsurl}[1]{\href{http://adsabs.harvard.edu/abs/#1}{ADS}}
\newcommand{\doiurl}[1]{\href{http://dx.doi.org/#1}{DOI}}

\newcommand{\eg}{e.g. {\ }}
\newcommand{\ie}{i.e. {\ }}

\newcommand{\fov}{field of view}
\newcommand{\kms}{~km~s$^{-1}$}

\renewcommand{\deg}{$^\circ$}
\newcommand{\Rsun}{\,R$_\odot$}
\newcommand{\Bsun}{$\overline{B_\odot}$} 
\begin{document}

\begin{article}

\begin{opening}

\title{The State of the White-Light Corona over the Minimum and Ascending Phases of Solar Cycle 25\,--\,Comparison with Past Cycles}

\author[addressref={aff1},corref,email={philippe.lamy@latmos.ipsl.fr}]{\inits{P. }\fnm{Philippe }\lnm{Lamy}\orcid{0000-0002-2104-2782}}
\author[addressref={aff1},email={hugo.gilardy@latmos.ipsl.fr}]{\inits{H. }\fnm{Hugo }\lnm{Gilardy}}

\address[id=aff1]{Laboratoire Atmosph\`eres, Milieux et Observations Spatiales, CNRS \& UVSQ, 11 Bd d'Alembert, 78280 Guyancourt, France}
      
\runningauthor{P. Lamy \& H. Gilardy}
\runningtitle{Coronal Activity in SC\,25}
 
%=====================================================================================================================================
\begin{abstract}
%=====================================================================================================================================
We report on the state of the corona over the minimum and ascending phases of Solar Cycle (SC) 25 on the basis of the temporal evolutions of its radiance and of the properties of coronal mass ejections (CMEs) as determined from white-light observations performed by the SOHO/LASCO-C2 coronagraph.
These evolutions are further compared with those determined during the past two Solar Cycles using the same methods.
The integrated radiance of the K-corona and the occurrence rate of CMEs closely track the indices/proxies of solar activity, prominently the total magnetic field for the radiance and the radio flux for the CMEs, all undergoing a steep increase during the ascending phase of SC\,25.
This increase is much steeper than anticipated on the basis of the predicted quasi similarity between SC\,25 and SC\,24, and is confirmed by the recent evolution of the  sunspot number. 
The radiance reached  the same base level during the minima of SC\,24 and 25, but the latitudinal extent of the streamer belt differed, being flatter during the latter minimum and in fact more similar to that of the minimum of SC\,23. 
Phasing the descending branches of SC\,23 and 24 led to a duration of SC\,24 of 11.0 years, similar to that given by the sunspot number. 
In contrast, the base level of the occurrence rate of CMEs during the minimum of SC\,25 was significantly larger than during the two previous minima.
The southern hemisphere is conspicuously more active than the northern one in agreement with several predictions and the current evolution of the hemispheric sunspot numbers.
In particular, the occurrence rate of the subset of CMEs with known mass, their mass rate, and the number of CMEs with speeds larger than 350\kms in the southern hemisphere exceeds by far the respective values in the northern hemisphere.
The mean apparent width of CMEs and the number of halo CMEs remains at relatively large, constant levels throughout the early phase of SC\,25 implying the persistence of weak total pressure in the heliosphere.
These results and particularly the perspective of a corona more active than anticipated are extremely promising  for the forthcoming observations by both {\it Solar Orbiter} and {\it Parker Solar Probe}.

\end{abstract}

\end{opening}

%===========================================================================================
\section{Introduction}
%===========================================================================================
The white-light images obtained with the ``LASCO-C2'' {\it Large-Angle Spectrometric COronagraph}  (\cite{Brueckner1995}) of the {\it Solar and Heliospheric Observatory} (SOHO: \cite{Domingo1995}) have allowed an unprecedented continuous coverage of the activity of the solar corona, starting in 1996 and still on-going.
A first analysis of the temporal evolution of the white-light corona over the first 18.5 years (1996.0\,--\,2014.5) based on these images was performed by \cite{Barlyaeva2015}.
They showed that the K-corona tracks solar activity at all time scales up to the solar cycle, including mid-term quasi-periodicities (also known as quasi-biennial oscillations or QBOs).
Among the various indices and proxies that they considered, the strongest correlation of the integrated coronal radiance was found with the total magnetic field. 
\cite{Lamy2014} compared in detail the solar minima of Solar Cycles (SC) 23 and 24 and found a 24\,\% decrease of the integrated radiance of the latter minima, but noted a very different behaviour of the northern and southern hemispheres with decreases of 17\,\% and 29\,\%, respectively. 
Phasing the ascending branches of the radiance of SC\,23 and 24, they estimated the duration of SC\,23 at 12 years and 3 months.
The eruptive activity of the corona over 23 years (1996.0\,--\,2019.0) was investigated by \cite{Lamy2019} in their review of coronal mass ejections (CMEs) that compared their properties reported by five catalogs, one manual (CDAW) and four automated (ARTEMIS, CACTus, SEEDS, and CORIMP).
They found that the occurrence and mass rates track the indices/proxies of solar activity likewise the radiance of the corona, but that the strongest correlation was with the radio flux F10.7. 
However the correlation coefficients were different during the two solar cycles, implying that the CME rates were relatively larger during SC\,24 than during SC\,23. 
Another striking feature of SC\,24 was the significant deficit in both occurrence and mass rates of CMEs in the southern hemisphere in comparison with the northern one.

With new LASCO-C2 data now extending to the beginning of 2022, thus covering the complete minimum phase of SC\,25 and its ascending phase, we are in position to extend our past analysis.
We are particularly interested in characterizing these phases and comparing them with those of the past solar cycles on the basis of the evolution of the radiance of the K-corona and the CME activity.
Another valuable aspect of describing and quantifying the present state of the corona consists in presenting the context for the on-going solar space missions and particularly, the instruments imaging the corona: the {\it Wide Field Imager for Solar Probe} (WISPR; \cite{Vourlidas2016}) on {\it Parker Solar Probe} (PSP; \cite{Fox2016}), the Metis coronagraph \citep{Antonucci2020} and the {\it Solar Orbiter Heliospheric Imager} (SoloHI; \cite{Howard2020}) on {\it Solar Orbiter} (SOLO; \cite{Muller2020}).

The present article makes use of images of the radiance of the K-corona and of the ARTEMIS-II catalog part of the LASCO-C2 Legacy Archive\footnote{\url{http://idoc-lasco-c2-archive.ias.u-psud.fr}} hosted at the Integrated Data and Operation Center (formerly MEDOC) of Institut d'Astrophysique Spatiale.
It is organized as follows:
In Section~\ref{Sec:Op}, we briefly summarize the operations of SOHO and of LASCO and its performance.
Section~\ref{Sec:Kcor} is devoted to the analysis of the state of the K-corona and Section~\ref{Sec:CME} to the properties of CMEs.
In Section~\ref{Sec:discussion}, we discuss our results in the broader context of solar activity and the predictions for Solar Cycle 25, and we conclude in Section~\ref{Sec:conclusion}.

%===========================================================================================
\section{SOHO and LASCO Operations}
     \label{Sec:Op} 
%===========================================================================================
As far as LASCO is concerned, the operation of SOHO during the past years was nominal.
SOHO continues to be periodically (every three months) rolled by 180\deg\ so that solar north periodically alternates between up and down on the LASCO images in the instrument reference frame.
The SOHO attitude is such that its reference orientation is the perpendicular to the ecliptic plane causing the projected direction of the solar rotational axis to oscillate between $\pm$\,7\deg\,15' around the ``vertical'' direction on the LASCO images.

\begin{figure}[htbp!]
	\centering
	\includegraphics[width=0.89\textwidth]{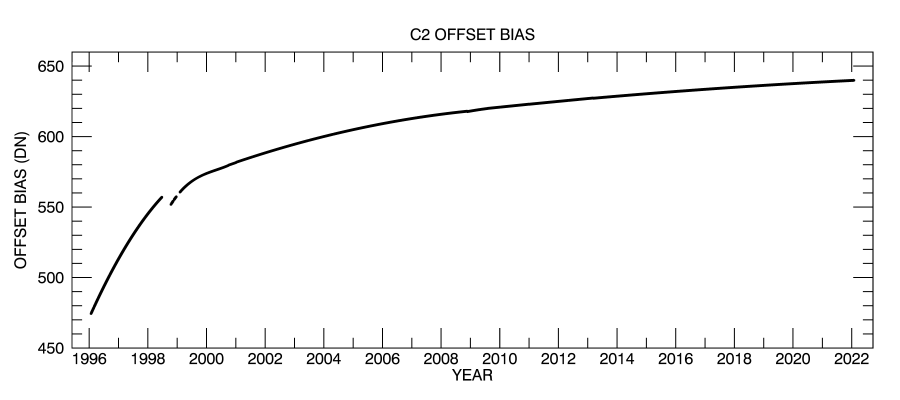}
	\includegraphics[width=\textwidth]{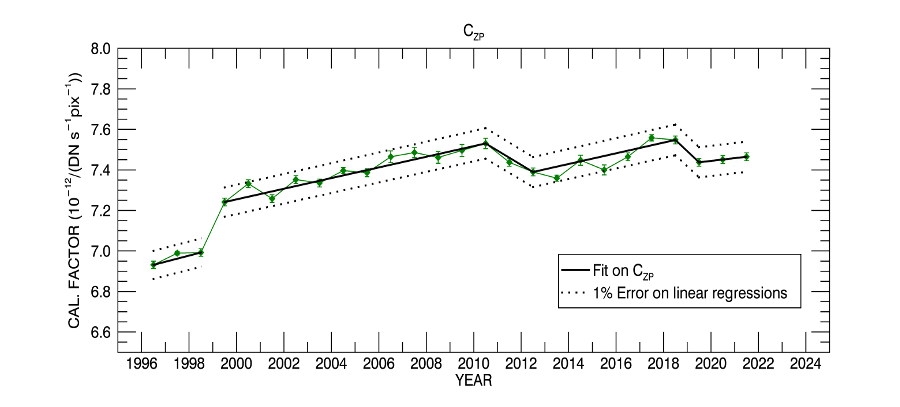}
	\caption{Upper panel: temporal variation of the offset bias of the LASCO-C2 camera.
	Lower panel: temporal variation of the LASCO-C2 calibration factor for the orange filter in units of $10^{-12}\,\overline{{\rm B}_\odot}/({\rm DN}\,{\rm sec}^{-1}\,{\rm pix}^{-1})$.}
	\label{Fig:calib}
\end{figure}

The in-flight performances of C2 slowly evolved as a consequence of the aging of the instrument.
As an example, Figure~\ref{Fig:calib} illustrates the evolutions of the offset bias of the CCD detector and of the calibration factor updated to the end of 2021.
The rapid evolution of the offset bias that prevailed during the first years of operation continues to level off so that the change during the last few years was very limited.
The calibration factor for the orange filter derived from thousands measurements of stars present in the C2 \fov\ (\cite{Llebaria2006}; \cite{Gardes2013}) exhibits a general increasing trend translating the continuous decline of the sensitivity of the C2 detector at a rate of typically 0.3\,\% per year. 
Apart from the sudden jump in 1999 linked to the ``hibernation'' interval when SOHO lost its pointing, the two decreases that took place in (2011\,--\,2013) and (2019\,--\,2021) have probably their origin in the electronics of the instrument.
This led us to introduce six linear functions to represent the temporal evolution of the factor. 
Inside each of the six regimes, the deviations of the measurements from the linear fits do not exceed 1\,\%.

\begin{figure}[htpb!]
	\centering
	\includegraphics[width=\textwidth]{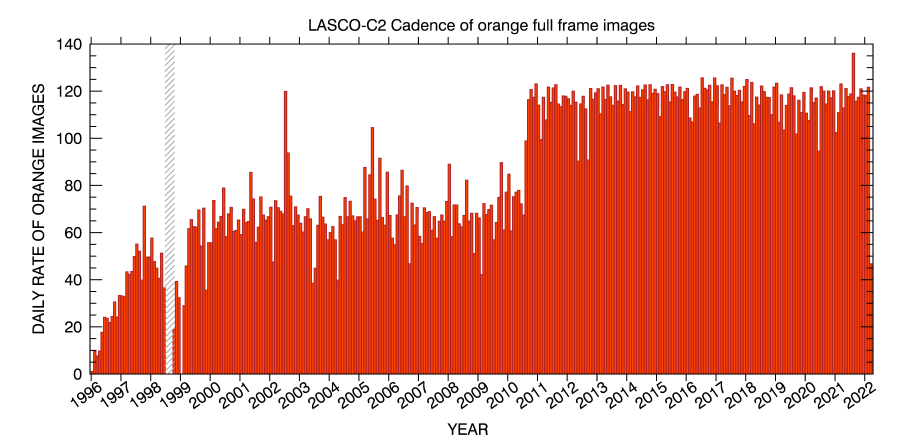}
	\includegraphics[width=0.97\textwidth]{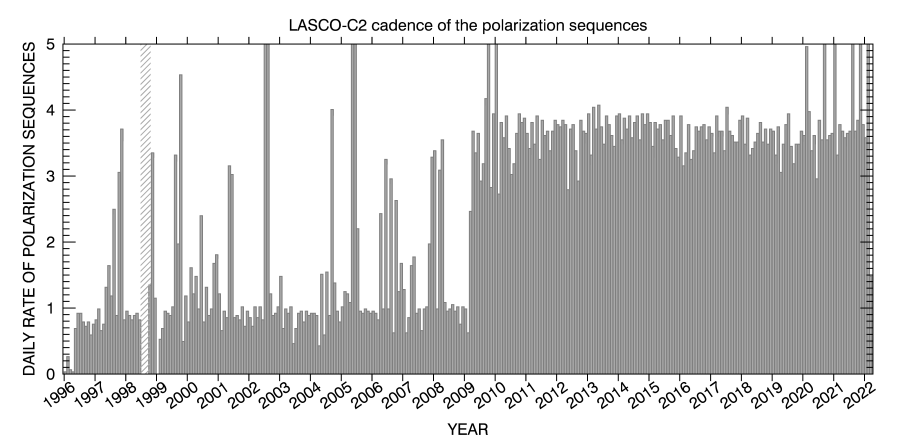}
	\caption{Monthly averaged daily rates of the routine radiance images (upper panel) and of the polarization sequences (lower panel) obtained with the LASCO-C2 coronagraph with the orange filter.
	The out-of-scale values of the latter rate correspond to high cadence polarization campaigns. Those that took place during the last two years were performed during successive perihelion passages of the {\it Parker Solar Probe}.}
	\label{Fig:cadence}
\end{figure}

The cadences of the routine full-frame images of 1024 $\times$ 1024 pixels used to detect and characterize the CMEs and of the polarization sequences generating the images of the radiance of the K-corona at the format of 512 $\times$ 512 pixels,  all taken with the orange filter, have remained at about the same past levels.
As illustrated in Figure~\ref{Fig:cadence}, the monthly averaged daily rates amounted to $\approx$\,120 and $\approx$\,4 for the full-frame images and the polarization sequences, respectively.

%========================================================================================
\section{Characterization of the K-Corona}
\label{Sec:Kcor} 
%========================================================================================
The polarization sequences were processed following the method developed by \cite{Lamy2020} to produce calibrated images of the polarized radiance $pB$, of the radiance of the K-corona $B_K$, and of the electron density.
As these three quantities behave in a similar way, our present analysis is limited to $B_K$.

%------------------------------------
\subsection{Structure of the K-corona} 
\label{Sec:K_struc}
\begin{figure}[htpb!]
	\centering
	\includegraphics[width=\textwidth]{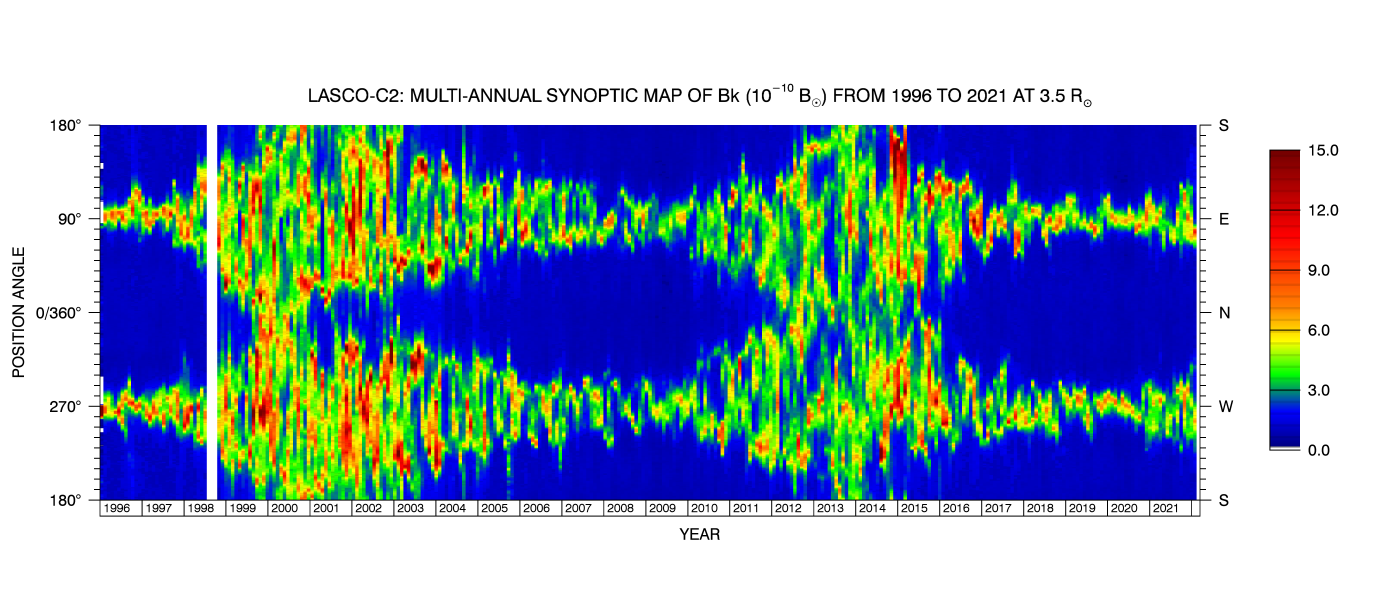}
	\caption{Multi-annual synoptic map of the radiance of the K-corona expressed in units of $10^{-10}$\,\Bsun\ at an elongation of 3.5\Rsun.} 
	\label{Fig:smap}
\end{figure}

%------------------------------------
An overview of the temporal evolution of the global structure of the corona is given by the multi-annual synoptic map of the radiance of the K-corona over 26.2 years (1996\,–-\,2022.2) at an elongations of 3.5\Rsun\ (Figure~\ref{Fig:smap}).
The difference between the maxima of SC\,23 and of SC\,24 is visually overwhelming and has already been addressed in many past publications. 
Equally striking are the differences between the three minima in both strength and latitude extent of the streamer belt; this will be further explored below.
A final noteworthy feature is the fast development of the activity in the rising phase of SC\,25 in comparison with the two past cycles, but this will become more obvious when considering the temporal evolution of the global radiance which is the topics of the next section.

%---------------------------------------------------
\subsection{Temporal Evolution of the Radiance of the K-corona}
\label{Sec:K_evolution} 
\begin{figure}[htpb!]
	\centering
	\includegraphics[width=\textwidth]{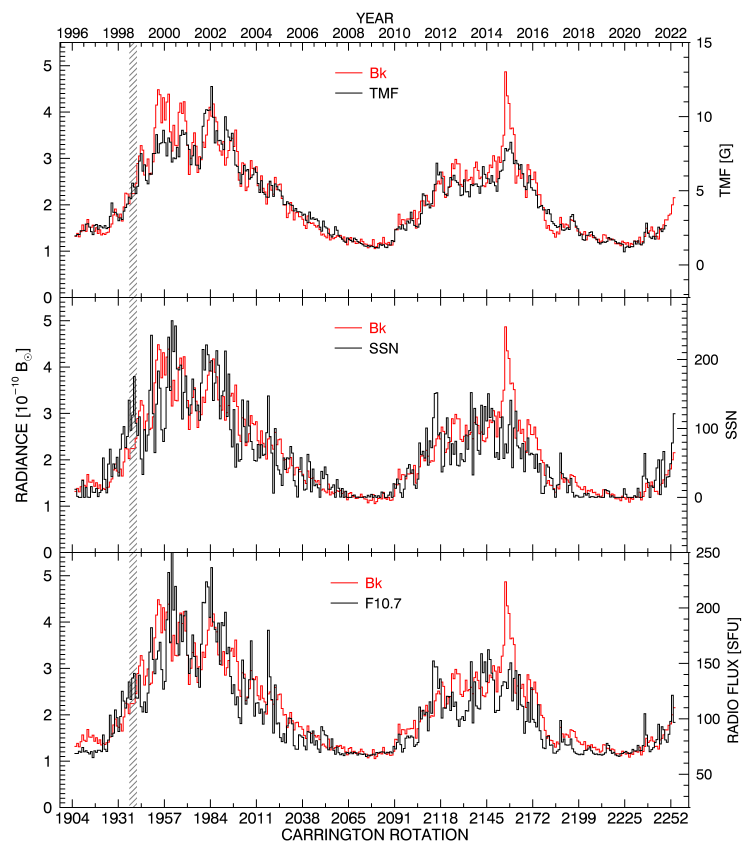}
	\caption{Comparison of the temporal variation of the global radiance of the K-corona integrated from 2.7 to 5.5 \Rsun\ with those of the total magnetic field (TMF expressed in units of Gauss), the sunspot number (SSN), and the radio flux at 10.7~cm (F10.7) expressed in Solar Flux Unit ($10^{-22}\,{\rm W}\,{\rm m}^{-2}\,{\rm Hz}^{-1})$).
	All quantities are averaged over the Carrington rotations.} 
	\label{Fig:Bk_compare}
\end{figure}

%---------------------------------------------------
Following our past works (\eg \cite{Barlyaeva2015}), the radiance of the K-corona was globally integrated in an annular region extending from 2.7 to 5.5\Rsun\ and the individual values were averaged over the Carrington rotations.
These authors compared the temporal variation with six indices and proxies of solar activity, but we limit the present comparison to the three most relevant:  
Sunspot Number (SSN)\footnote{http://sidc.oma.be/silso/datafiles}, Total Photospheric Magnetic Flux (TMF)\footnote{Courtesy Y.-M.~Wang}, and Decimetric Radio Flux at 10.7~cm (F10.7)\footnote{http://www.ngdc.noaa.gov/stp/space-weather/solar-data/solar-features/solar-radio/}.
The first two are photospheric indices, whereas F10.7 combines chromospheric and coronal activities.
Due to an equipment failure at the Wilcox Solar Observatory, the monthly reductions of the photospheric maps have been delayed and, as a consequence, the TMF data are presently available only until CR 2249 (21 October 2021).
Figure~\ref{Fig:Bk_compare} confirms the conclusion of \cite{Barlyaeva2015} that the integrated radiance tracks the indices of solar activity, the highest correlation being with the TMF followed by the F10.7 and SSN. 
Unlike these last two indices, TMF and $B_K$ agree in many small scale variations and further, in the remarkable increase that took place at the end of 2014 and persisted to early 2015.
It resulted from an unusual configuration of the magnetic field following the emergence of the large sunspot complex AR 12192 in October 2014 as analyzed by \cite{Sheeley2015}.
It caused the coronal plasma to be trapped at low latitudes and prevented CMEs from erupting.
This process inflated a bulge in the corona creating an anomalous surge of brightness extensively analyzed by \cite{Lamy2017}.

The last few years of the descending phases of SC\,23 and 24 are remarkably similar reaching the same base level during the following minima as will be discussed in the next section.
Whereas we correlated the ascending branches of SC\,23 and 24 to set the duration of SC\,23 at 12 years and three months, we correlated their descending branches to set the duration of SC\,24 at 11.0 years (\ie the canonical duration of a solar cycle), in agreement with the value determined by the WDC-SILSO data center.

%Figure~\ref{Fig:Bk_compare} displays the global radiance of the K-corona from January 1996 to mid-March 2022 together with the total magnetic field (TMF), the sunspot number (SSN) and the radio flux at 10.7~cm (F10.7).
%The general trend follows very well the variation of these standard proxies of solar activity, thus demonstrating the close correlation of the modulation of the global radiance in phase with the solar cycle.

\begin{figure}[htpb!]
	\centering
	\includegraphics[width=\textwidth]{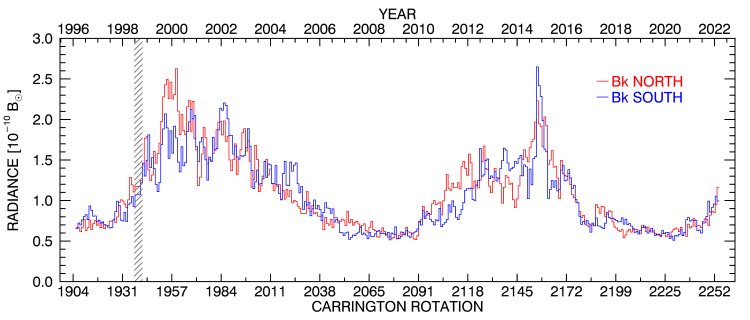}
	\caption{Comparison of the temporal variation of the global radiance of the K-corona integrated from 2.7 to 5.5 \Rsun\ and averaged over the Carrington rotations in the northern and southern hemispheres.} 
	\label{Fig:Bk_NS}
\end{figure}

Figure~\ref{Fig:Bk_NS} displays the temporal variation of the integrated radiance separately in the northern and southern hemispheres.
Whereas marked differences were present during SC\,24 particularly during its rising and maximum phases, they tended to disappear thereafter and the two variations are remarkably similar during the minimum and ascending phases of SC\,25.

%-------------------------------------------------------------------
\subsection{Comparison of the Minima of Solar Cycles 23, 24, and 25}
\label{Sec:minimaK} 
\begin{figure}[htpb!]
	\centering
	\includegraphics[width=\textwidth]{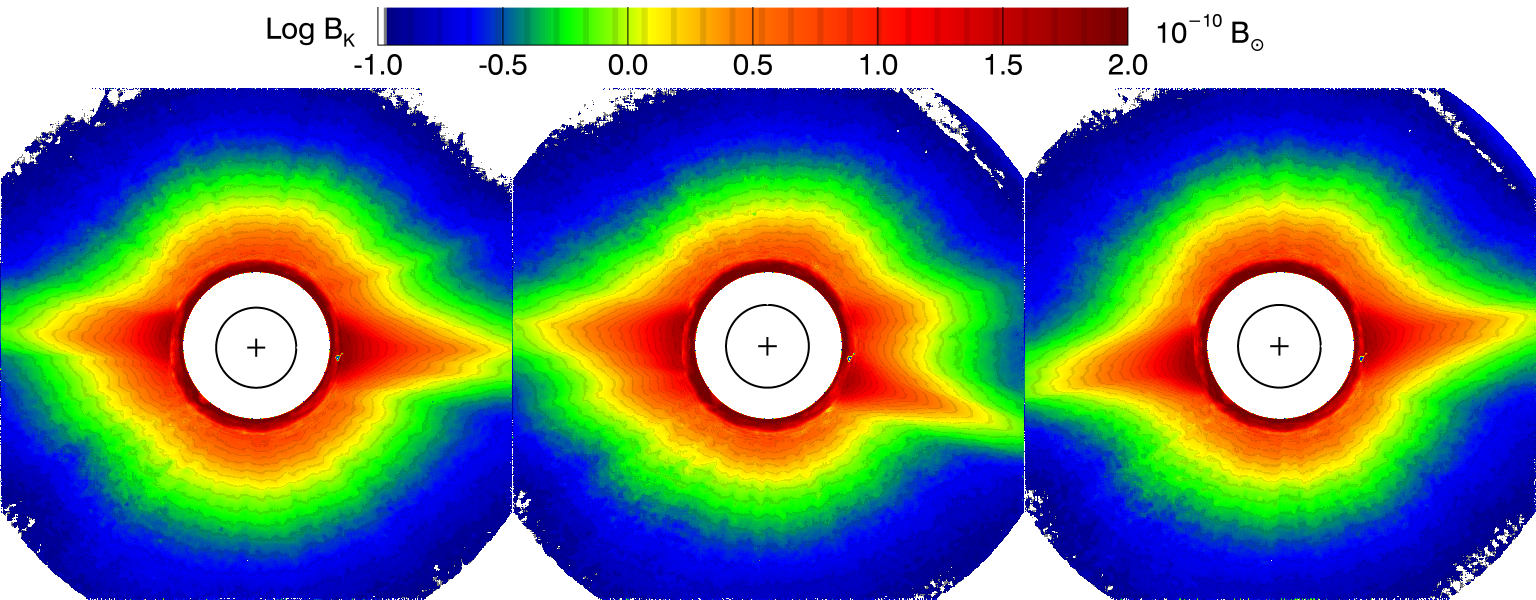}
	\caption{Three images of the K-corona obtained with the LASCO-C2 coronagraph at three consecutive minima of solar activity: SC\,23 (left panel), SC\,24 (middle panel), and SC\,25 (right panel). 
  The radiance is expressed in units of $10^{-10}$\,\Bsun\ and its logarithm is coded according to the color bar.} 
	\label{Fig:3minimaK}
\end{figure}

%-------------------------------------------------------------------
Our detailed analysis of the three minima starts with the comparison of three typical images of the K-corona as illustrated in Figure~\ref{Fig:3minimaK}.
The most striking difference already alluded to in Section~\ref{Sec:K_struc} concerns its structure during the prolonged anomalous minimum of SC\,24.
As analyzed by \cite{Lamy2014}, the observed latitudinal extent of the streamer belt remained large, consistent with a large tilt angle of the heliocentric current sheet \citep{Manoharan2012} and a low polar field \citep{Petrie2013}.
In contrast, the minima of SC\,23 and 25 are characterized by the quasi similar structure of a flat streamer belt corresponding to the simple ``dipole'' geometry of the large-scale magnetic field of solar minima. 
Conversely, the coronal holes reached their usual low latitudes of approximately $15^o$ to $20^o$ suggesting that the polar field was restored to its ``nominal'' level during the SC\,25 minimum.

\begin{figure}[htpb!]
	\centering
	\includegraphics[width=\textwidth]{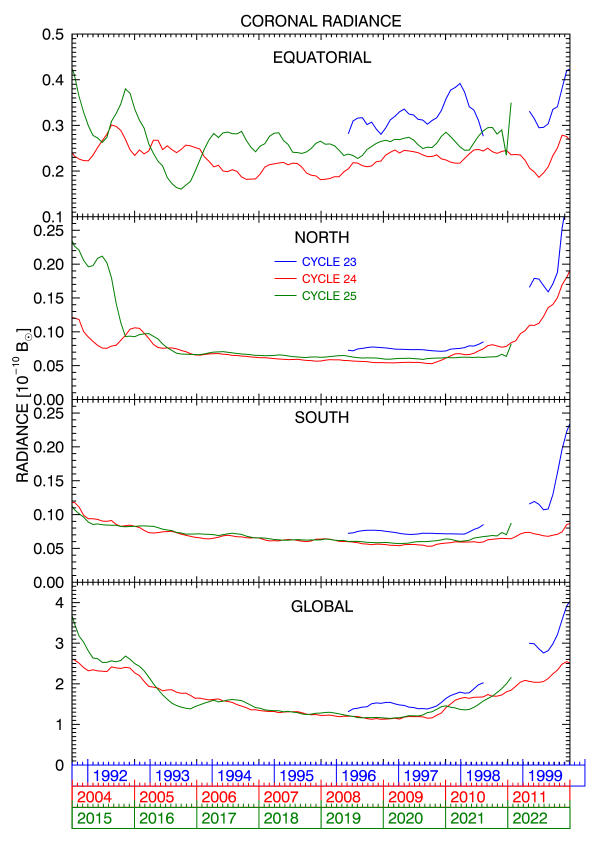}
	\caption{Results of phasing the temporal variations of the radiance of the K-corona during the minima of Solar Cyles 23, 24, and 25.
	The radiance is globally integrated from 2.7 to 5.5\Rsun\ and then in different sectors: southern, northern, and equatorial.
	The gaps in the blue curves (SC\,23) correspond to the loss of SOHO.} 
	\label{Fig:phasing}
\end{figure}

The comparison is pursued by considering the detailed temporal evolution of the integrated radiance of the K-corona between 2.7 and 5.5\Rsun\ over eight years centered on the three minima conveniently named first (SC\,23), second (SC\,24), and third (SC\,25).
In addition, different sectors were considered as defined by \cite{Barlyaeva2015}: the northern and southern sectors are centered on the polar direction and the equatorial sector combines the east and west sectors centered on the equatorial direction; all of them have a full angular width of 30\deg.
Following their procedure, we shifted the time intervals in order to phase the minima according to the duration of the solar cycles as already determined: 12 years and 3 months for SC\,23 and 11.0 years for SC\,24 (Figure~\ref{Fig:phasing}).
The behaviours and the base levels of the radiance temporal profiles of the second and third minima are remarkably similar in the case of the global corona and of the southern and northern sectors with a very slight increase in the latter case, the third minima exceeding the second one by $\approx$10\,\%.
In contrast, the base level of the first minimum was significantly larger, exceeding that of the second one by $\approx$32\,\% as determined by \cite{Lamy2014}. 
The situation is completely different in the equatorial sector with much larger differences.
Averaging the local fluctuations, we estimated an increase of the base levels of $\approx$30\,\% between the first and third minima and up to $\approx$50\,\% between the first and second minima.
The onset of the ascending branch of SC\,25 took place in early May 2020 on the basis of the temporal profiles of the global coronal and the south sector, but clearly the north sector is lagging by at least 1.5 year with a hint of a take-off in October 2021.

%========================================================================================
\section{Coronal Mass Ejections over the Minimum and Ascending Phases of Solar Cycle 25}
\label{Sec:CME} 
%========================================================================================
This analysis makes use of the ARTEMIS-II catalog which relies on the detection of CMEs on synoptic maps based on their morphological appearance (\cite{Boursier2009}; \cite{Floyd2013}).
It lists CMEs detected since June 1996 with the following parameters: time of detection at 3\,\Rsun, central apparent latitude, angular width, and intensity.
The intensity of a CME is calculated by integrating its radiance on the synoptic maps at 3\,\Rsun\ and subtracting the local coronal background. 
It does not strictly corresponds to the total radiance as recorded on the images, but it offers a valuable estimate of the CME strength \citep{Lamy2019}.
For a large fraction of the CMEs ($\approx$\,60 \%), the catalog further lists three different velocities (``propagation'', ``global'', and ``median''), mass, and kinetic energy.
We consider below the last two velocities obtained by cross-correlating the detected CMEs on the original synoptic maps at 3 and 5.5\,\Rsun.
A global cross-correlation yields the global velocity whereas a line by line cross-correlation produces a distribution of velocities whose median value is taken as the median velocity.
The global velocity gives more weight to the front and central parts whereas the median velocity gives an equal weight to every angular section of the CME.
As a consequence, the former is systematically larger than the latter.
For the time interval considered in this study, from 6 June 1996 to 7 February 2022, the ARTEMIS-II (hereafter abbreviated to ARTEMIS for simplicity) catalog lists a total of 42,165 CMEs, of which 23,885 have their velocities, mass, and kinetic energy determined.

%---------------------------------------------------
\subsection{Occurrence and Mass}
\label{Sec:occur} 
\begin{figure}[htpb!]
	\centering
	\includegraphics[width=\textwidth]{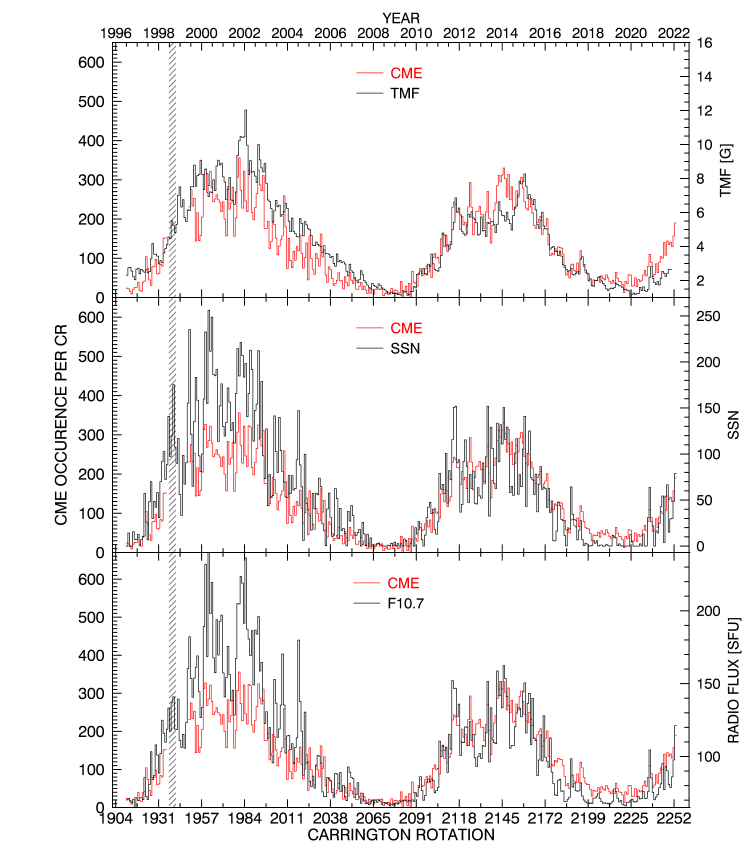}
	\caption{Temporal variation of the CME occurrence rate per Carrington rotation compared with those of the total magnetic field (TMF), the sunspot number (SSN), and the radio flux at 10.7~cm (F10.7) expressed in Solar Flux Unit ($10^{-22}\,{\rm W}\,{\rm m}^{-2}\,{\rm Hz}^{-1})$)}.
	\label{Fig:rate}
\end{figure}

%---------------------------------------------------
We first consider the temporal variation of the occurrence rate of the whole set of CMEs corrected for the LASCO duty cycle as described by \cite{Lamy2019} and calculated per Carrington rotation.
Likewise the case of the radiance of the corona, a comparison was performed with the selected indices and proxies of solar activity: SSN, TMF, and F10.7 whose variations were adequately scaled and shifted so as to best fit that of the CMEs during SC\,24.
Indeed and as shown in Figure~\ref{Fig:rate}, it is not possible to perform the fits over both SC\,23 and 24 as the relationships were different during the two cycles.
In their review, \cite{Lamy2019} used SC\,23 as a reference, but we now favor SC\,24 because of its immediate proximity to SC\,25. 
Their conclusion obviously still holds: the CME occurrence rate was relatively larger during SC\,24 than during SC\,23 as also found by \cite{Gopal2015} using the sunspot number. 
The new data reveals that the occurrence rate started to diverge from the common trend of the indices and proxies of solar activity in the last years of the declining phase of SC\,24 implying an excess of CMEs with respect to the evolution of solar activity.
This divergence persisted during the minimum and ascending phases of SC\,25, particularly pronounced when comparing with the radio flux. 
Section~\ref{Sec:minimaCME} below analyzes in detail the situation of this minimum compared with the two past ones.
Regarding the situation of the ascending phase of SC\,25, two years after its minimum the CME occurrence rate has reached a high of 5.2 CME per day whereas it was only 3.7 CME per day two years after the minimum of SC\,24, a significant increase.

\begin{figure}[htpb!]
	\centering
	\includegraphics[width=\textwidth]{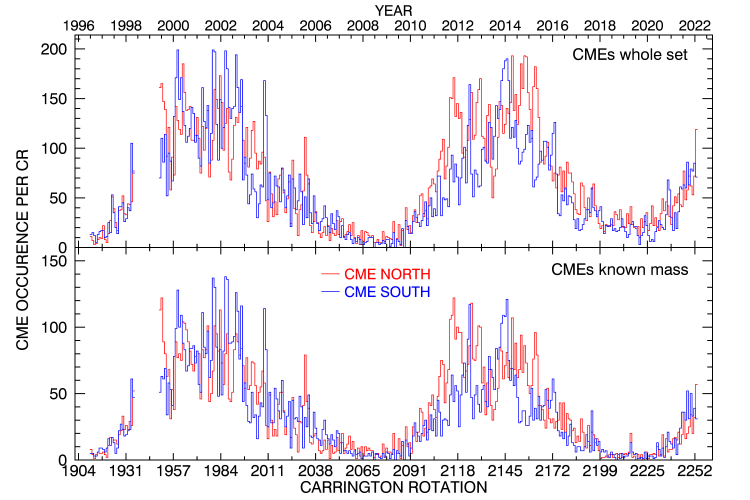}
	\caption{Temporal variation of the CME occurrence rates per Carrington rotation in the northern and southern hemispheres.
	The upper panel corresponds to the whole set of CMEs and the lower one is restricted to the CMEs with known mass.}
	\label{Fig:rate_NS}
\end{figure}

We next consider the temporal variation of the occurrence rate of the whole set of CMEs separately in the northern and southern hemispheres (upper panel of Figure~\ref{Fig:rate_NS}).
There was a slight excess of southern CMEs by a mere 2.3\,\% during SC\,23 followed by a vigorous reversal during SC\,24 with an excess of northern CMEs by an astonishing 28\,\% (see also Table~\ref{Table:stat}).
The same trend persisted during the early years of SC\,25, but at a much reduced level of 11\,\% until early February 2022 and this resulted from the combined effect of a larger base level during the minimum ($\approx$\,0.9 versus $\approx$\,0.7 CME per day) and a steeper rate of occurrence in the northern versus the southern hemispheres. 
However, it appears that the southern rate is catching up during the latest CRs.

\begin{figure}[htpb!]
	\centering
	\includegraphics[width=\textwidth]{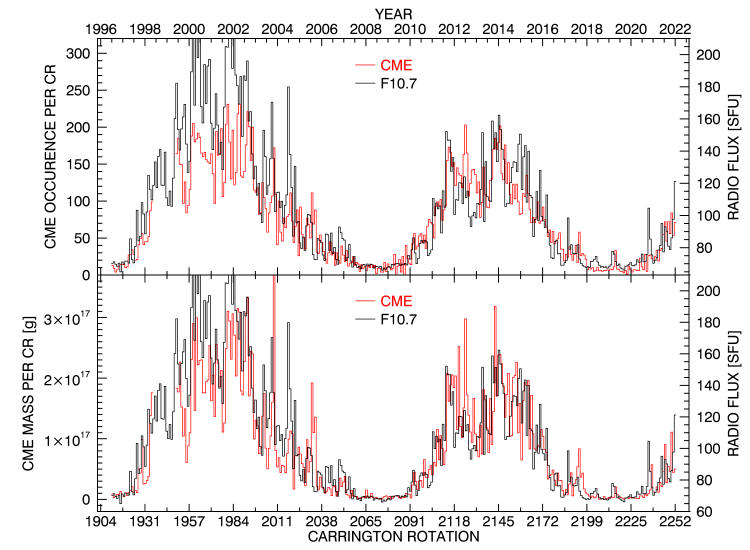}
	\caption{Temporal variations of the occurrence rate of CMEs with known mass (upper panel) and of the CME mass rate (lower panel) per Carrington rotation compared with that of the F10.7 radio flux expressed in Solar Flux Unit ($10^{-22}\,{\rm W}\,{\rm m}^{-2}\,{\rm Hz}^{-1}$).}
	\label{Fig:mass}
\end{figure}

The set of CMEs with known mass presently amounts to 57\,\% of the global set, the missing 43\,\% corresponding to faint CMES for which complete characterization could not be achieved (\eg lack of detection on three consecutive synoptic maps).
The temporal evolution of the occurrence rate of these CMEs is displayed in the upper panel of Figure~\ref{Fig:mass} together with that of the radio flux F10.7.
Their correlation during SC\,24 is better than in the case of the whole set of CMEs except during two restricted time intervals, from late 2012 to early 2013 and from mid-2014 to mid-2015.
Unlike the case of the whole set of CMEs, the occurrence rate of CMEs with known mass closely tracks the radio flux during the minimum and ascending phases of SC\,24. 
This implies that the relative excess found for the whole set prominently results from the relative overabundance of faint CMEs.

We next consider the temporal variation of the occurrence rate of the set of CMEs with known mass separately in the northern and southern hemispheres (lower panel of Figure~\ref{Fig:rate_NS} and Table~\ref{Table:stat}).
There was a slight excess of northern CMEs of 4\,\% during SC\,23 followed by a vigorous reversal during SC\,24 with an excess of northern CMEs of 30\,\%.
This is very much consistent with the whole set of CMEs so that the two sets of CMEs followed a similar trend during SC\,23 and 24, namely a quasi equilibrium of northern and southern CMEs during SC\,23 and a large excess of northern CMEs of approximately 30\,\% during SC\,24.
The situation is entirely different during the early years of SC\,25 with an excess of the whole set of northern CMEs of 11\,\% and an excess of southern CMEs with known mass of 15\,\%.
As pointed out above, this results from the relative overabundance of faint CMEs in the northern hemisphere, but this is probably a temporary effect as the southern rate is catching up during the latest CRs.
Ultimately, we expect a situation opposite to that of SC\,24, that is a large excess of southern CMEs during SC\,25.

Turning to the mass of CMEs accumulated per Carrington rotation, the lower panel of Figure~\ref{Fig:mass} shows that it better tracks the radio flux than the occurrence rate.
This is especially true over SC\,24 (note the similar local drop centered on 2013.0) and during the minimum and ascending phases of SC\,24. 

%In a broad stroke, the temporal evolution of the intensity (determined for the global set) behaves very much like the mass rate when compared with the radio flux so that the above remarks apply as well. 
%This is not a surprise as \cite{Lamy2019} showed that, for the set of CMEs with known mass, intensity and mass are highly correlated. 
%They exploited this property to estimate the ``missing'' mass -- that is the mass of the CMEs for which it could not be determined -- and found that it amounts to $\approx$14\,\% of the measured mass, thus confirming that these ``missing'' CMEs are indeed faint.

%-------------------------------------------------------------------
\subsection{Comparison of the Minima of Solar Cycles 23, 24, and 25}
\label{Sec:minimaCME} 
\begin{figure}[htpb!]
	\centering
	\includegraphics[width=\textwidth]{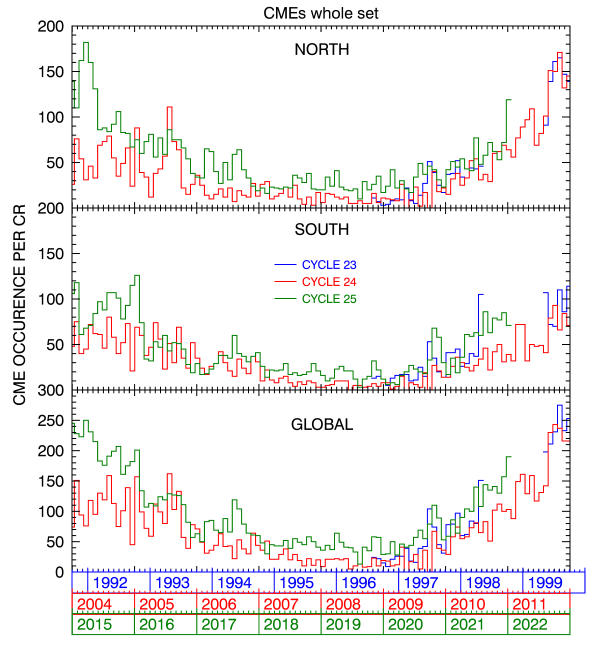}
	\caption{Results of phasing the temporal variations of the occurrence rate of the whole set of CMEs during the minima of Solar Cyles 23, 24, and 25.
	The rates were calculated globally and separately in the southern and northern hemispheres.
	The gaps in the blue curves (SC\,23) correspond to the loss of SOHO.} 
	\label{Fig:phasingCMEall}
\end{figure}

\begin{figure}[htpb!]
	\centering
	\includegraphics[width=\textwidth]{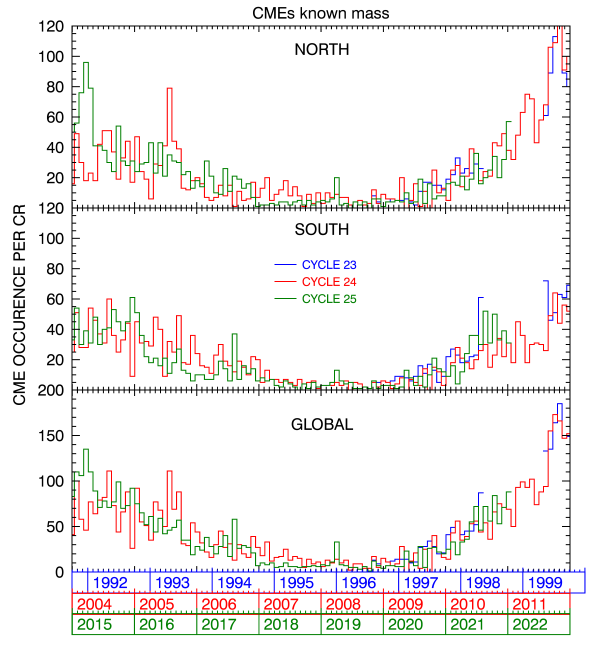}
	\caption{Results of phasing the temporal variations of the occurrence rate of the set of CMEs with known mass during the minima of Solar Cyles 23, 24, and 25.
	The rates were calculated globally and separately in the southern and northern hemispheres.
	The gaps in the blue curves (SC\,23) correspond to the loss of SOHO).} 
	\label{Fig:phasingCMEmass}
\end{figure}

%-------------------------------------------------------------------
ikewise the case of the radiance of the K-corona, we consider the detailed temporal evolution of the occurrence rates of the CMEs over eight years centered on the three minima.
The time intervals were shifted in order to phase the minima according to the determined duration of the solar cycles.  
Figure~\ref{Fig:phasingCMEall} illustrates the case of the whole set of CMEs.
A first striking feature is the quasi systematically enhanced rates of CMEs during the last minimum in comparison with the previous two, with only a few local exceptions. 
As a consequence, the base level of the global (N+S) population during the minimum of SC\,25 amounts to $\approx$\,1.5 CME per day, significantly higher than during the two previous minima, $\approx$\,0.85 and $\approx$\,0.67 CME per day during the minima of SC\,23 and 24, respectively.
A second striking feature is the steeper increase of the rate during the ascending phase of SC\,25 in comparison with the two past similar phases, this effect being entirely produced by the southern CMEs. 
Figure~\ref{Fig:phasingCMEmass} illustrates the case of the set of CMEs with known mass.
Unlike the whole set of CMEs, the variations of the occurrence rates over the eight years are approximately similar implying a common base level of $\approx$\,0.37 (N+S) CME per day.
A steeper increase of southern CMEs during the ascending phase of SC\,25 is suggested, but less pronounced than in the case of the whole set of CMEs.
The combination of these results confirms that the CMEs in excess number during the minimum of SC\,25 are prominently faint events, but that the overall population of CMEs contributes to the steep increase of the occurrence rate in the southern hemisphere.

%---------------------------------------------------
\subsection{Angular Width}
\label{Sec:width} 
\begin{figure}[!htpb]
\centering
\includegraphics[width=\textwidth]{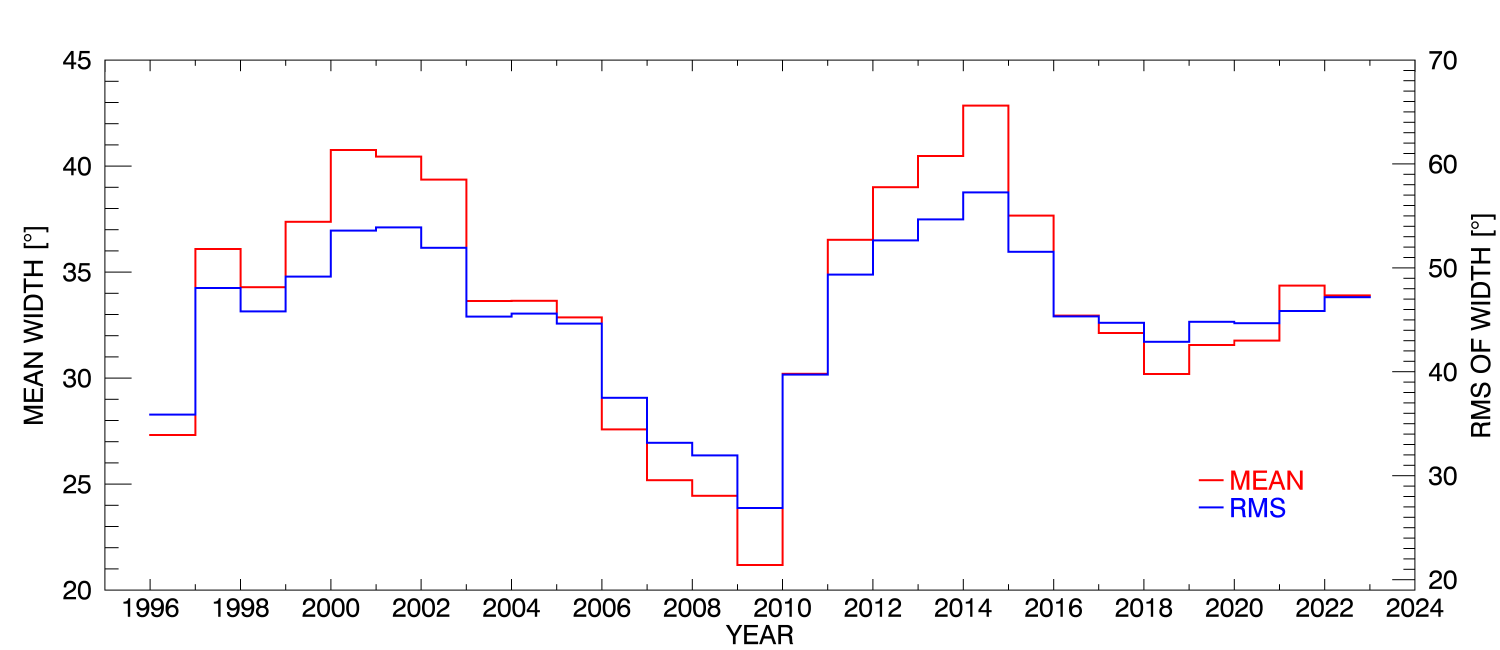}
\caption{Temporal variation of the annualized mean and root-mean-squared values of the apparent angular width of CMEs narrower than 180\deg.} 
\label{Fig:yearly_aw_180}
\end{figure}

%---------------------------------------------------
Figure~\ref{Fig:yearly_aw_180} displays the temporal variation of the annualized mean and root-mean-squared values of the apparent angular width $W$ of CMEs narrower than 180\deg, thus extending a similar figure produced by \cite{Lamy2019}.
It now appears that the mean width was significantly larger during the minimum of SC\,25 than during the minimum of SC\,24, approximately 43\deg\ versus 25\deg.
In addition, the usual trend of increasing widths as solar activity develops has not yet materialized during the rising phase of SC\,25.

\begin{figure}[!htpb]
\centering
\includegraphics[width=\textwidth]{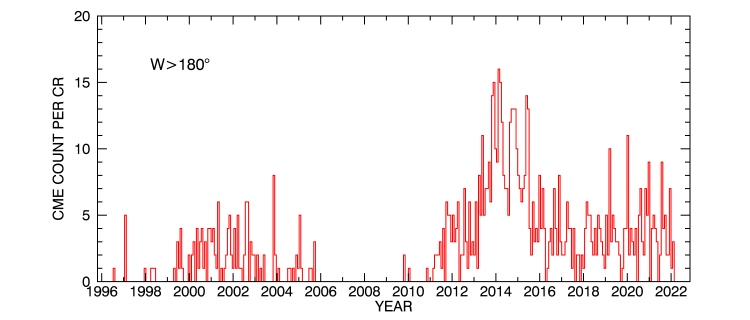}
\includegraphics[width=\textwidth]{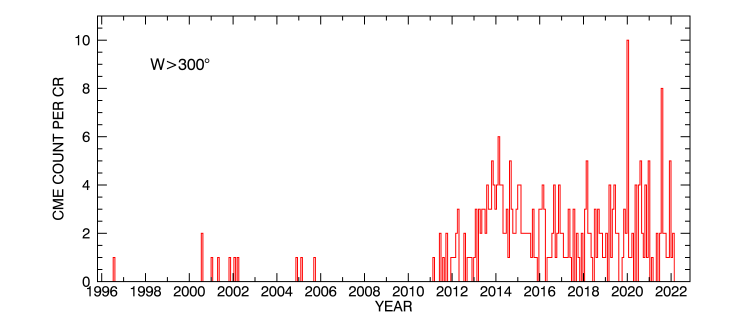}
\caption{Monthly occurrence rates of CMEs with width $>$180\deg\ (upper panel) and with width $>$300\deg\ (lower panel).}
\label{Fig:halo}
\end{figure}

Regarding the distribution of angular widths, the number of CMEs during the last three years was relatively small so that these CMEs do not affect our past result based on the previous 23 years, prominently controlled by the two maxima. 
As a reminder, the distribution of angular width follows an exponential law characterized by a mean value of 42\deg\ and a constant slope of -0.0107 in the range 40\deg\ to 300\deg.
The restriction to CMEs of known mass marginally changes these values to 44\deg\ and -0.0106, respectively.
There is a clear turnover in the distributions at $\approx$\,300\deg\ used to define the regime of halo CMEs.
We introduce two intervals of width, $>$180\deg\ and $>$300\deg, to account for partial halos \citep{Gopal2003}, and Figure~\ref{Fig:halo} displays the monthly occurrence rate of ARTEMIS CMEs in these two intervals. 
The much larger rates during SC\,24 compared with SC\,23 was already highlighted by \cite{Lamy2019}, but we now see that this trend persists during the ascending phase of SC\,25.
Even more striking is the presence of full and partial halo CMEs throughout the minimum of SC\,25 whereas they were merely absent during the past two minima.

%---------------------------------------------------
\subsection{Apparent Latitude}
\label{Sec:lat} 
\begin{figure}[htpb!]
	\centering
	\includegraphics[width=\textwidth]{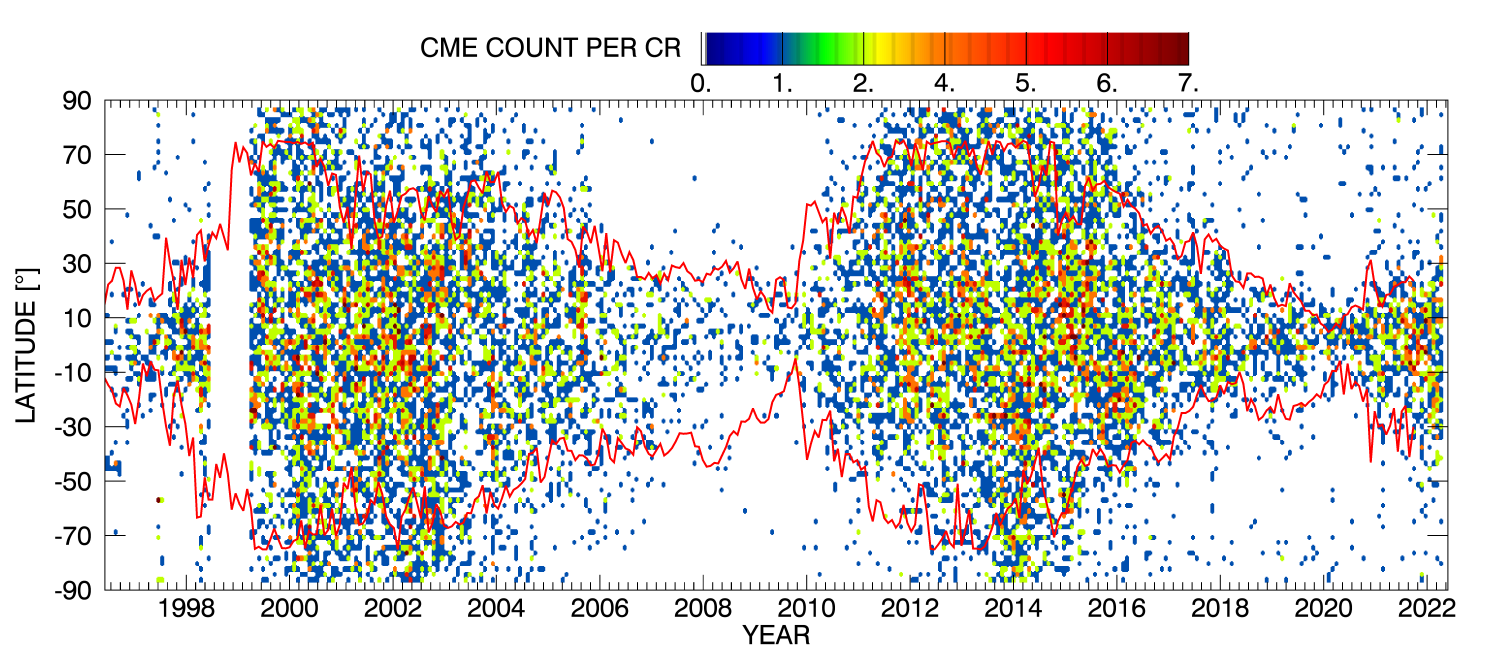}
	\includegraphics[width=\textwidth]{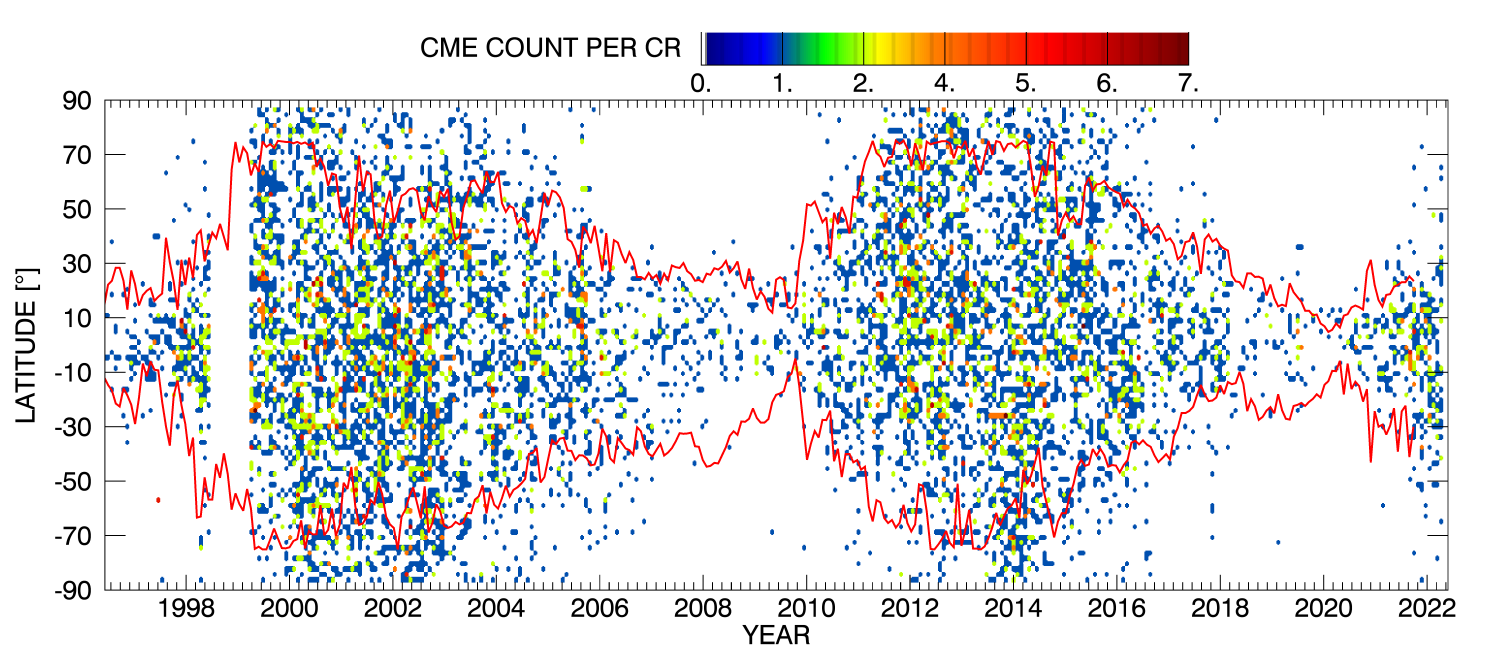}
	\caption{Heliolatitudinal distributions of the whole set of CMEs (upper panel) and the set of CMEs with known mass (lower panel).
	The counts are calculated per Carrington rotation and per latitude interval of 2\deg\ and are color coded according to the color bar.
	The red lines correspond to the tilt angle of the heliospheric current sheet in the northern and southern hemispheres.} 
	\label{Fig:CMElat}
\end{figure}

\begin{figure}[htpb!]
	\centering
	\includegraphics[width=\textwidth]{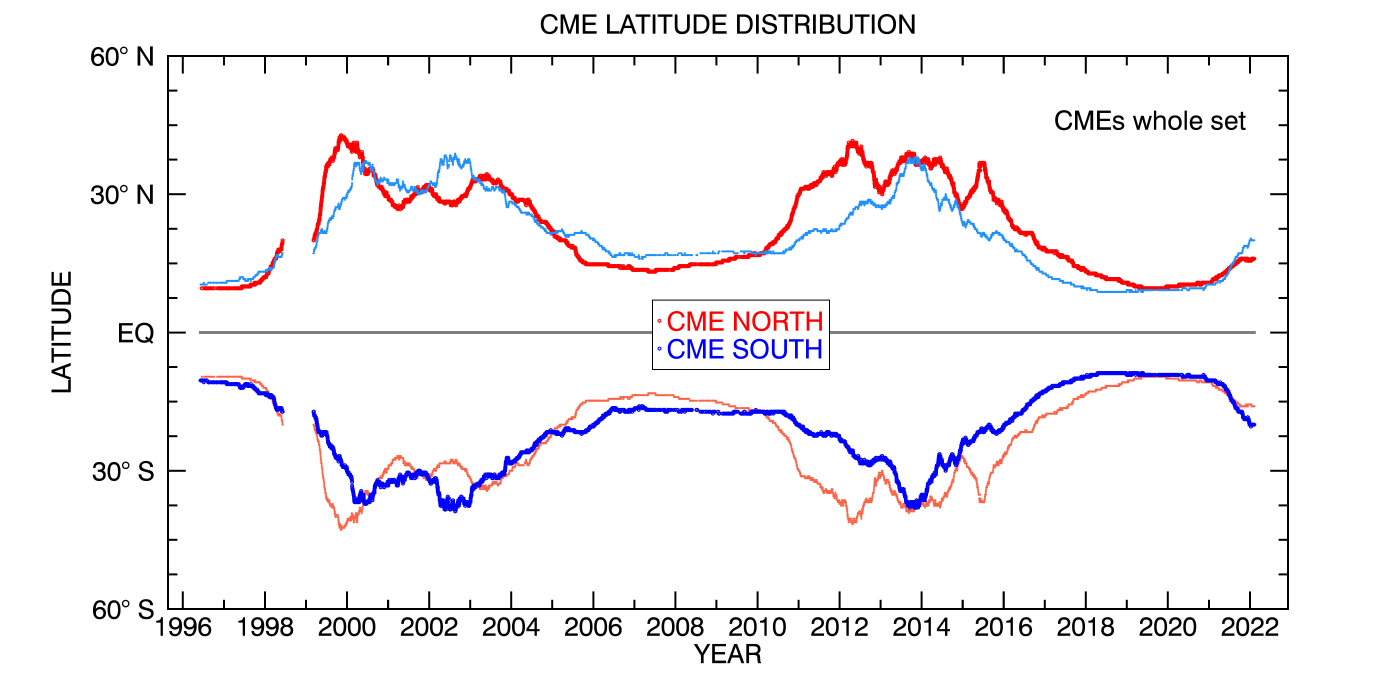}
	\includegraphics[width=\textwidth]{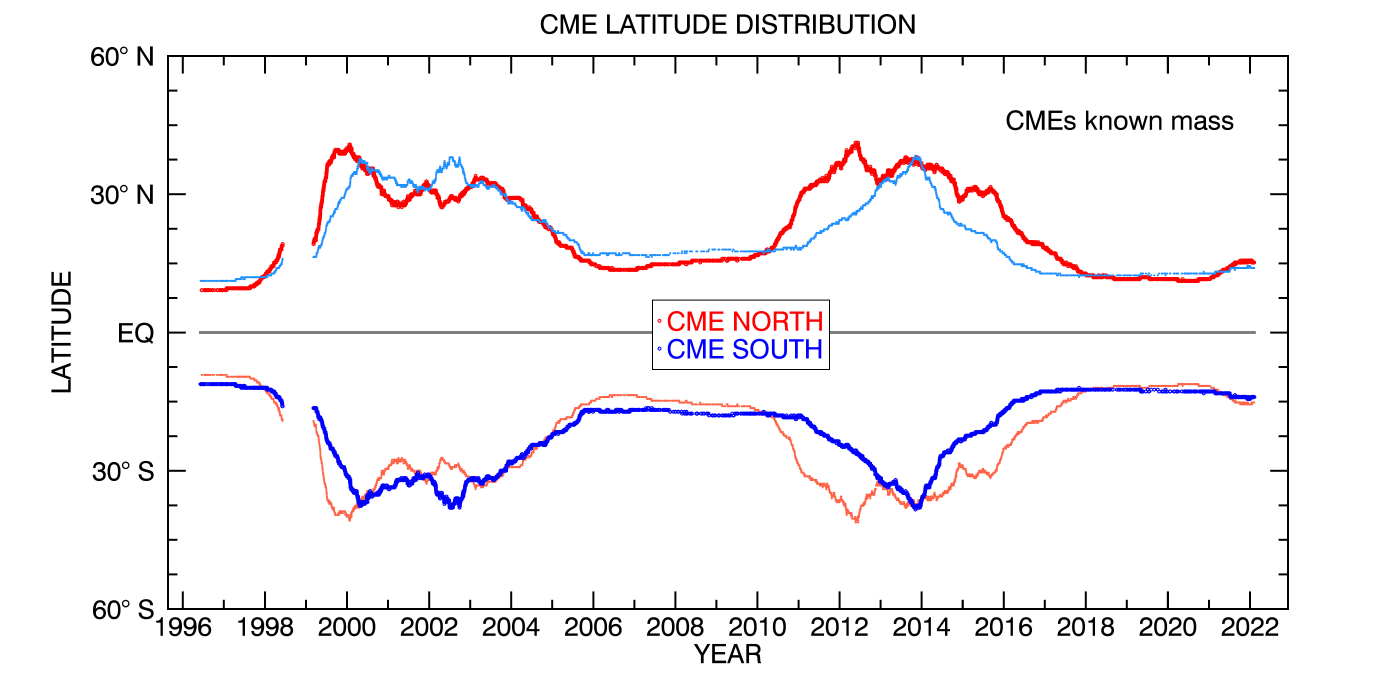}
	\caption{Temporal variation of the mean value per Carrington rotation of the CME apparent latitude separately in the northern (red line) and southern (blue line) hemispheres.
	The upper panel corresponds to the whole set of CMEs and the lower one is restricted to the CMES with known mass.
	The symmetric of the two curves (light red and light blue lines) are over-plotted to facilitate the comparison.} 
	\label{Fig:CMElat_mean}
\end{figure}

%---------------------------------------------------
The spatial distribution of apparent latitudes of CMEs is best perceived on the heliolatitudinal maps displayed in Figure~\ref{Fig:CMElat} where CMEs are counted in boxes defined by a Carrington rotation and a latitude interval of 2\deg.
We superimposed the evolution of the tilt angle of the heliospheric current sheet (HCS) provided by the Wilcox Solar Observatory\footnote{\url{http://wso.stanford.edu/Tilts.html}} using the ``classic'' potential field model as recommended.
The two maps corresponding to the whole set of CMEs and to the set of CMEs with known mass are highly consistent, implying that the restriction on mass does not induce any bias.
Concentrating on SC\,25, a noteworthy feature is the fact that the distribution of apparent latitudes remains well bounded by the tilt of the HCS which was not the case during the minimum of SC\,24.
A synthetic view is offered by Figure~\ref{Fig:CMElat_mean} which displays the evolution of the mean value per Carrington rotation of the apparent latitude separately in the northern and southern hemispheres. 
In the case of the whole set, there appears during the last six months a hint of an asymmetry between the two hemispheres with CMEs present at larger southern latitudes than at north latitudes in agreement with the tilt angle.
If confirmed in the near future, this trend would be consistent with the faster development of CME activity in the southern hemisphere pointed out in the above sections.

%---------------------------------------------------
\subsection{Kinematics}
\label{Sec:kine} 
\begin{figure}[htpb!]
	\centering
	\includegraphics[width=\textwidth]{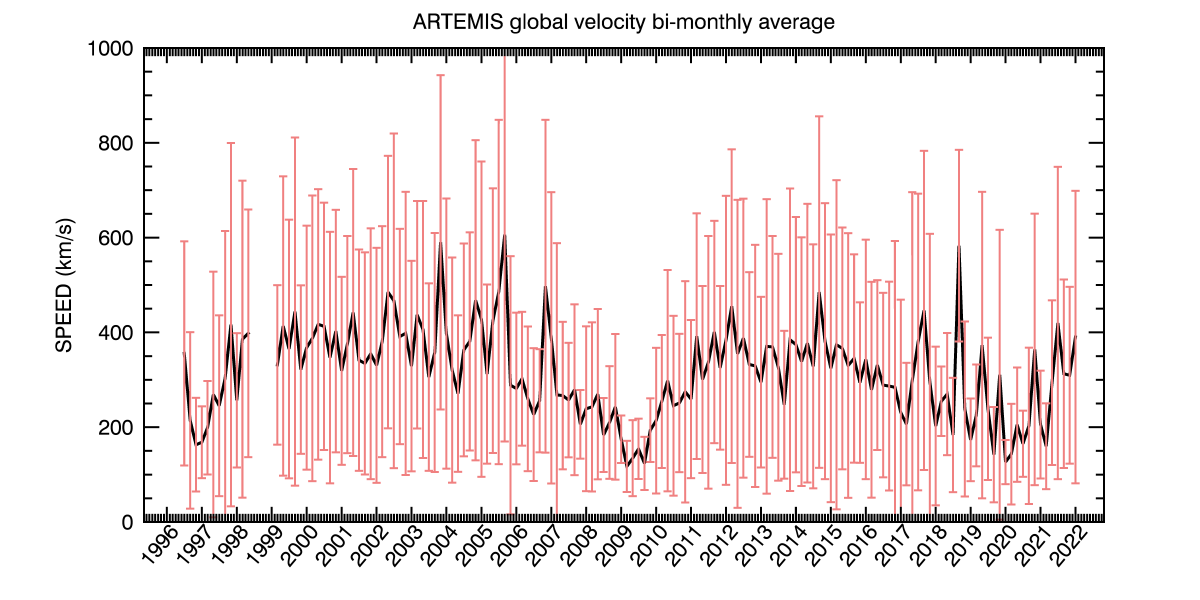}
	\includegraphics[width=\textwidth]{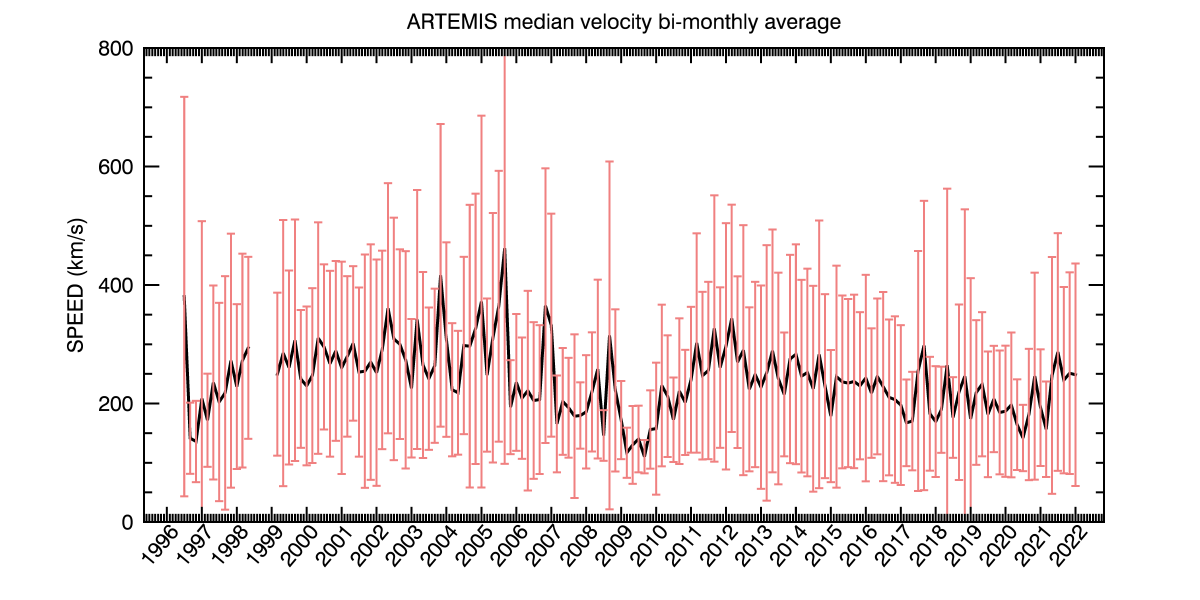}
	\caption{Temporal variation of the bi-monthly average values of the global (upper panel) and median (lower panel) speeds of CMEs derived from the ARTEMIS catalog.} 
	\label{Fig:speed}
\end{figure}

\begin{figure}[htpb!]
	\centering
	\includegraphics[width=\textwidth]{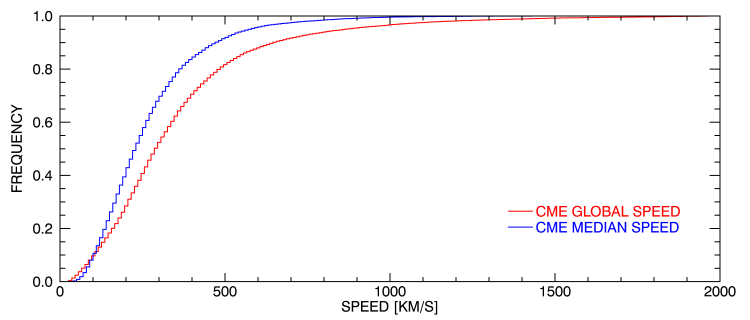}
	\caption{Cumulative distribution functions of the apparent global and median speeds of CMEs derived from the ARTEMIS catalog.} 
	\label{Fig:speed_cumul}
\end{figure}

%---------------------------------------------------
Figure~\ref{Fig:speed} displays the annual variation of the mean and standard deviation values of the global and median speed distributions reported by the ARTEMIS catalog.
We used bi-monthly average values in order to smooth the short-scale fluctuations while preserving the detail of the variation during the solar cycles.
The trends of i) the speeds tracking solar activity and ii) larger speeds during SC\,23 compared with the weaker SC\,24 have already been underlined (\eg \cite{Lamy2019}). 
The new salient feature concerns the last two minima of solar activity: whereas both speeds experienced a drastic reduction during the minimum of SC\,24, it was less pronounced during that of SC\,25 and furthermore, nearly absent in the case of the median speed.
The cumulative distributions of the two speeds calculated until 7 February 2022  are displayed in Fig.~\ref{Fig:speed_cumul}.
As expected, the spread of the global speeds is larger than that of the median speeds and the median values reach 280 and 220\kms\ for the global and median speeds, respectively.

\begin{figure}[htpb!]
	\centering
	\includegraphics[width=\textwidth]{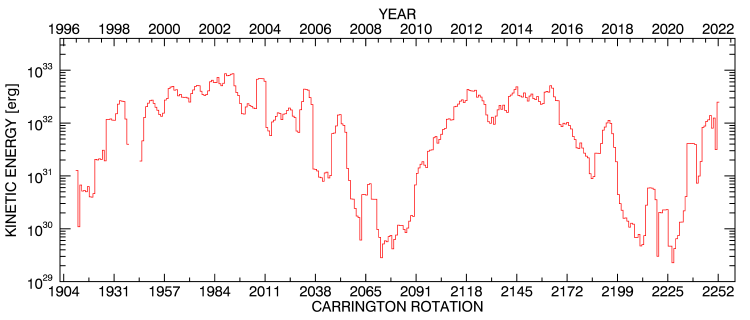}
	\caption{Temporal evolution of the kinetic energy of CMEs per Carrington rotation.} 
	\label{Fig:ke}
\end{figure}

Fig.~\ref{Fig:ke} displays the temporal evolution of the kinetic energy of CMEs per Carrington rotation.
Curiously, the minimum of SC\,25 witnessed episodes of rather energetic CMEs and the following rising phase is particularly abrupt when compared with that of the previous cycle. 

%---------------------------------------------------
\subsection{Summary Statistics of Coronal Mass Ejections}
\label{Sec:stat} 
%---------------------------------------------------
Table~\ref{Table:stat} presents an updated version of Table~9 of \cite{Lamy2019} synthesizing the statistical properties of the whole set of CMEs reported by the ARTEMIS catalog until 7 February 2022.
The results for SC\,23 remain unchanged whereas those for SC\,24 were completed to the end of the cycle rounded to 31 December 2019; those for SC\,25 are naturally limited to its minimum and ascending phases. 
The last column sums up the above results and gives an overview of the properties of CMEs during the past 26 years. 

SC\,23 featured a relatively well-balanced activity in the two hemispheres, the only exception being a modest 10\,\% difference concerning those CMEs with width larger then 30\deg.
This was no longer the case during SC\,24 as the northern hemisphere was much more active than the southern one.
For instance, the occurrence rates of northern CMEs exceed the southern ones by 28\,\% and 30\,\% for the whole set of CMEs for that of CMEs with known mass, respectively.
The same situation prevailed for the other properties with unbalances ranging from 22\,\% to 37\,\% and culminating to a factor of 21 in the case of halo CMEs.
The first two years of SC\,25 offer a different picture.
The census of the whole set of CMEs exhibits a 12\,\% surplus of northern CMEs whereas the opposite situation prevails for the set of CMEs with known mass with a 15\,\% surplus of southern CMEs.
This implies that the northern hemisphere ejects a relatively large number of faint CMEs whose mass could not be determined by our procedure.
This is confirmed by the statistics on the mass for which the southern hemisphere systematically outperforms the northern one, and it further stands out by a larger population of fast CMEs with speeds $>$350\kms, the count of southern ones exceeding the northern one by an astonishing 51\,\%.

\begin{table}
\centering
\caption{Statistical properties of the LASCO CMES listed in the ARTEMIS catalog until 7 February 2022.}
\vspace{0.2cm}
\begin{tabular}{ccccc}
\noalign{\smallskip}\hline\noalign{\smallskip}
\multicolumn{5}{c}{ALL CMEs} \\
%\noalign{\smallskip}\Xhline{2\arrayrulewidth}\noalign{\smallskip}
\noalign{\smallskip}\hline\noalign{\smallskip}
Count 		& SC23 	& SC24 	& SC25 	& SC24+SC23+SC25 \\
\noalign{\smallskip}\hline\noalign{\smallskip}
%\noalign{\smallskip}\Xhline{2\arrayrulewidth}\noalign{\smallskip}

Total count                          	& 20194 & 19732 & 2415 & 42341 \\
Total count (north)                   &  9985 & 11092 & 1273 & 22350 \\
Total count (south)                   & 10209 &  8640 & 1142 & 19991 \\

Angular width $\leq 30^o$             & 11340 & 10835 & 1397 & 24572 \\
Angular width $\leq 30^o$ (north)     &  5764 &  5944 & 754 & 12462 \\
Angular width $\leq 30^o$ (south)     &  5576 &  4891 & 643 & 11110 \\
Angular width $>30^o$                 &  8854 &  8897 & 1018 & 18769 \\
Angular width $>30^o$ (north)         &  4221 &  5148 &  519 & 9888 \\
Angular width $>30^o$ (south)         &  4633 &  3749 &  499 & 8881 \\
Angular width $>300^o$ (halos)        &    11 &   231 &   59 & 301 \\

Speed $\leq 350$\kms                  &  7397 &  6594 & 668 & 14659 \\
Speed $\leq 350$\kms (north)          &  3659 &  3767 &  328 & 7754 \\
Speed $\leq 350$\kms (south)          &  3738 &  2827 &  340 & 6905 \\
Speed $>350$\kms                     	&  5443 &  3599 &  269 & 9311 \\
Speed $>350$\kms (north)              &  2646 &  1981 &  107 & 4734 \\
Speed $>350$\kms (south)              &  2797 &  1618 &  162 & 4577 \\

%\noalign{\smallskip}\Xhline{2\arrayrulewidth}\noalign{\smallskip}
\noalign{\smallskip}\hline\noalign{\smallskip}
\multicolumn{5}{c}{CMEs WITH KNOWN MASS} \\
\noalign{\smallskip}\hline\noalign{\smallskip}
%\noalign{\smallskip}\Xhline{2\arrayrulewidth}\noalign{\smallskip}
Count/Mass 		& SC23 	& SC24 	& SC25 	& SC24+SC23+SC25 \\
\noalign{\smallskip}\hline\noalign{\smallskip}
%\noalign{\smallskip}\Xhline{2\arrayrulewidth}\noalign{\smallskip}

Total count                  	&12840         & 10193     & 937         & 23970\\
Total count (north)           & 6305         & 5748         & 435         & 12488\\
Total count (south)           & 6535         &  4445     & 502         & 11482\\
\noalign{\smallskip}
Total mass (g)                & 1.5E+19 & 1.1E+19 & 7.8E+17 & 2.7E+19 \\
Total mass (g) (north)        & 7.6E+18 & 6.9E+18 & 3.4E+17 & 1.5E+19 \\
Total mass (g) (south)        & 7.7E+18 & 4.3E+18 & 4.4E+17 & 1.2E+19 \\

Mean mass (g)                 & 1.2E+15 & 1.1E+15 & 8.3E+14 & 1.1E+15 \\
Mean mass (g) (north)         & 1.2E+15 & 1.2E+15 & 7.7E+14 & 1.2E+15 \\
Mean mass (g) (south)         & 1.2E+15 & 9.7E+14 & 8.8E+14 & 1.1E+15 \\

Median mass (g)             	& 3.0E+14 & 2.6E+14 & 2.5E+14 & 3.3E+14 \\
Median mass (g) (north)     	& 3.0E+14 & 2.6E+14 & 2.4E+14 & 2.8E+14 \\
Median mass (g) (south)     	& 3.0E+14 & 2.5E+14 & 2.7E+14 & 2.9E+14 \\

\noalign{\smallskip}\hline\noalign{\smallskip}
%\noalign{\smallskip}\Xhline{2\arrayrulewidth}\noalign{\smallskip}
\end{tabular}
\label{Table:stat}
\end{table}

%============================================================================================
\section{Discussion}
\label{Sec:discussion}
%============================================================================================
The prediction of amplitude and timing of SC\,25 has been the subject of many articles using a variety of techniques, and compilations may be found in \cite{Courtillot2021} covering the time interval (2011\,--\,2019) extended to 2021 by \cite{Burud2021} and \cite{Javaraiah2022}.
It is obviously beyond the scope of the present article to discuss this copious literature, but we may attempt to identify a general trend and note that the most recent publications converge to a strength of SC\,25 comparable to SC\,24 or slightly larger \citep{Kumar2021}. 
A few of them (\eg \cite{Burud2021}; \cite{Chowdhury2021}; \cite{Courtillot2021}) however concluded on a slightly weaker SC\,25 and suggested that it will witness the beginning of the upcoming Gleissberg cycle.
The forecast consensus of the NOAA/NASA co-chaired, international panel of 9 December 2019 concluded on a peak in July 2025 ($\pm$\,8 months), with a smoothed sunspot number of 115, hence similar to SC\,24.
However, the ascending phase of SC\,25 appears much steeper than this prediction\footnote{https:// www. swpc. noaa. Gov/ news/solar-cycle-25-forecast-update}.
In fact, the SNN monthly values of December 2021 (67.6), January (54) and February (59.7) 2022 are well above the forecast values of 26.6, 29.0, and 31.5, respectively.
%This pronounced difference by a factor of approximately two is confirmed by the F10.7 index.

The question of which solar hemisphere would be dominant was considered by a dozen articles recently reviewed by \cite{Javaraiah2022}. 
The consensus goes towards a marked north–south asymmetry with activity dominant in the southern hemisphere.
Particularly interesting are the converging results on the peak amplitudes in the northern and southern hemispheres using different methods: 66 and 83, respectively according to \cite{Pishkalo2021}, 64.3 and 83.8, respectively according to \cite{Gopal2022}. 
The present hemispheric sunspot numbers\footnote{http://sidc.oma.be/silso/datafiles} go in that direction with an excess of southern sunspots, the turnover having occurred at the onset of the rising phase of SC\,25.

The temporal evolution of the integrated radiance of the K-corona is in excellent agreement with the current steep increase of solar activity as it closely tracks the selected indices and proxies (Figure~\ref{Fig:Bk_compare}). 
The correlations with the TMF during the terminal part of the descending branch of SC\,24, and with the minimum and rising phases of SC\,25 are particularly impressive.
Furthermore, Figure~\ref{Fig:Bk_NS} reveals that the radiance in the southern hemisphere started to exceed the northern one during the past year.

We are aware of only one work, that of \cite{Mostl2020}, attempting to predict the CME ejection rate for SC\,25 or more precisely, the ICME rate with direct implication for Parker Solar Probe in situ observations. 
They did so by linking the SSN to the observed ICME rates in SC\,23 and 24 with the list of \cite{Richardson2010} and their own ICME catalog.
They then determined linear relationships that they extended to SC\,25 using nearly extreme predictions of its strength. 
By construction, their temporal evolution of the ICMEs rate tracks the SSN, but the rates themselves are very low, ranging from 15 per year (1 per CR) to 23 per year (1.7 per CR) at the beginning of 2020, compared with $\approx$\,140 per CR detected by LASCO.
This inherently results from the fact that the number of ICMEs detected in situ are far less than that of CMEs detected on high cadence coronagraphic images.

The occurrence rate of CMEs with known mass and the mass rate per CR closely track the radio flux (Figure~\ref{Fig:mass}) and therefore the current steep increase of solar activity, very much like the radiance.
Regarding the hemispheric occurrence rate of CMEs, the global set indicates an excess of northern CMEs of 12\,\%.
In strong contrast, the occurrence rate of southern CMEs of known mass outperforms the northern one by 15\,\%; this percentage increases to 30\,\% when considering the mass rate per CR and even to 51\,\% for CMEs with speeds $>$350\kms (Table~\ref{Table:stat}).
Clearly, the southern hemisphere develops an overwhelming activity as predicted.

Finally, the persistent large number of halo CMEs following the trend already observed during SC\,24 in comparison with SC\,23 (Figure~\ref{Fig:halo}) is best explained by the weak total pressure in the heliosphere that prevailed after the anomalous minimum of SC\,24 and which facilitates the widening of CMEs so as to become halo more frequently (\cite{Gopal2020}; \cite{Gopal2022}).

%============================================================================================
\section{Conclusion}
\label{Sec:conclusion}
%============================================================================================
In this article, we have presented the state of the white-light corona over the minimum and ascending phases of SC\,25 on the basis of the analysis of the temporal variation of its radiance $B_K$ and its CME production rates and properties.
Both closely track the indices/proxies of solar activity, prominently the total magnetic field for $B_K$ and the radio flux for the CMEs.
Their evolution confirms the steep increase of the rising phase of SC\,25, much steeper than anticipated on the basis of predicted sunspot numbers quasi similar  to those of SC\,24. 
This is obviously of utmost interest for the forthcoming observations of the corona by SOLO and PSP.
We highlight below our most significant results.
\begin{itemize}
\item 
The global radiance of the corona integrated between 2.7 and 5.5\Rsun\ reached  the same base level during the minima of SC\,24 and 25, but the latitudinal extent of the streamer belt differed, being flatter during the latter minimum and in fact more similar to that of the minimum of SC\,23.
\item 
The correlation of the descending branches of SC\,23 and 24 led to a duration of SC\,24 of 11.0 years similar to that given by the sunspot number. 
\item 
The occurrence rate of the global set of CMEs, when adjusted to the variation of the indices/proxies during SC\,24, started to diverge at the onset of the minimum of SC\,25, the base level being significantly larger than during the previous minima.
\item 
This is not the case of the occurrence rate of the set of CMEs with known mass which closely sticks to the variation of the indices/proxies, and so does the rate of CME mass per CR.
This implies that the excess CMEs are faint, modestly contributing to the mass budget.
\item 
The southern hemisphere appears significantly more active than the northern one in agreement with several predictions and the current evolution of the hemispheric sunspot numbers.
In particular, the occurrence rate of the set of CMEs with known mass, the total mass of CMEs, and the number of CMEs with speeds larger than 350\kms in the southern hemisphere exceeds by far the respective values in the northern hemisphere.
\item 
The mean apparent width of CMEs during the minimum of SC\,25 did not drop to the low values reached during the previous minimum and it remains at a nearly constant level throughout the early phase of SC\,25.
A similar trend is observed in the case of the occurrence rate of halo CMES.
\end{itemize}

%============================================================================================
\vspace{\baselineskip} 
\noindent
{\footnotesize\bf Acknowledgments}
We thank Y.-M. Wang for providing the total magnetic field data.
The LASCO-C2 project at the Laboratoire Atmosph\`eres, Milieux et Observations Spatiales is funded by the Centre National d'Etudes Spatiales (CNES).
LASCO was built by a consortium of the Naval Research Laboratory, USA, the Laboratoire d'Astrophysique de Marseille (formerly Laboratoire d'Astronomie Spatiale), France, the Max-Planck-Institut f\"ur Sonnensystemforschung (formerly Max Planck Institute f\"ur Aeronomie), Germany, and the School of Physics and Astronomy, University of Birmingham, UK.
SOHO is a project of international cooperation between ESA and NASA.

\vspace{\baselineskip} 
\noindent
{\footnotesize\bf Disclosure of Potential Conflicts of Interest} The authors declare that they have no conflicts of interest.
%===========================================================================================================================================================

%=================================================================================================
\clearpage 
\bibliographystyle{spr-mp-sola}
\bibliography{SC25_Biblio}                % name your BibTeX data base 
\nocite{*}
%==================================================================================================

%==================================================================================================
%\appendix

%\section*{Detailed Description of the restoration of the Stray Light}
%\label{Sec:appendix}

%==================================================================================================

\end{article}

\end{document}